# Image-Based RKPM for Accessing Failure Mechanisms in Composite Materials

Yanran Wang [1], Yichun Tang [2], Jing Du [2], Mike Hillman [3], J. S. Chen [1*]

[1] Department of Structural Engineering, University of California San Diego, La Jolla, CA 92093-0085, USA

[2] Department of Mechanical Engineering, Pennsylvania State University, University Park, PA 16802, USA

[3] Karagozian & Case, Inc., Glendale, CA 91203, USA

**Abstract.** Stress distributions and the corresponding fracture patterns and evolutions in the microstructures strongly influence the load-carrying capabilities of composite structures. This work introduces an enhanced phase-field fracture model incorporating interface decohesion to simulate fracture propagation and interactions at material interfaces and within the constituents of composite microstructures. The proposed method employs an interface-modified reproducing kernel (IM-RK) approximation for handling cross-interface discontinuities constructed from image voxels and guided by Support Vector Machine (SVM) material classification. The numerical models are directly generated from X-ray microtomography image voxels, guided by SVM using voxel color code information. Additionally, a strain energy-based phase field variable is introduced, eliminating the need to solve coupled field problems. The effectiveness of this method is demonstrated in modeling crack growth both along interfaces and across matrix and inclusion domains and in predicting the corresponding structural-scale mechanical behavior in composite structures. Furthermore, the proposed method has been validated against experimentally observed crack patterns.

## 1 Introduction

In recent years, non-destructive imaging techniques like micro-X-ray computed tomography (micro-CT) have become powerful tools for examining the microstructure and internal deformation of composite materials [1-4]. Despite these advancements, the challenge of modeling microstructures persists due to their inherent geometrical complexity and heterogeneity, making mesh generation for mesh-based methods particularly computationally intensive.

[*] Corresponding author
e-mail address: jsc137@ucsd.edu (J. S. Chen)

This study uses the Support Vector Machine (SVM) algorithms for image segmentation to inform numerical model generation. SVM is a classification-based machine learning algorithm built on solid mathematical foundation and optimization frameworks [5-6]. SVM offers advantages over other methods as it generates a unique maximum-margined global hyperplane for separating training datasets, providing a global solution for data classification. Additionally, it is not sensitive to the underlying probabilistic distribution of the training dataset, ensuring high performance for limited, noisy, or imbalanced datasets [7]. The traditional binary SVM algorithm is adopted in this work for its effective applicability to the two-phase materials.

Numerical modeling of heterogeneous materials remains challenging for both mesh-based methods discretized with body-fitted discretization and meshfree methods formulated with smooth approximations. For the Finite Element Method (FEM), incomplete handling of discontinuities in mesh construction can lead to suboptimal convergence [8] and aligning meshes with interfaces is a non-trivial task for composites with complex microstructures and significant variations in constituent moduli. The Finite Cell Method is a high-order embedded domain technique [9] that provides a simple yet effective modification of traditional FEM to bypass the necessity of exhaust body-fitted meshing for geometrically and topologically complex microstructures. However, special numerical integration schemes are needed to differentiate between inside and outside of the physical domain. The meshfree methods utilize point-wise discretization instead of carefully constructed body-fitted meshes. However, meshfree methods such as element-free Galerkin (EFG) [10] and reproducing kernel particle method (RKPM) [11-13] typically suffer from Gibb's-like oscillation in the approximation when modeling weak continuities in composite materials, as their smooth approximation functions with overlapping local supports fail to capture gradient jump conditions across material interfaces [14]. This work introduces the Interface-Modified Reproducing Kernel Particle Method (IM-RKPM) [15], which leverages signed distance functions from SVM-classified micro-CT images to handle discontinuities across material interfaces without requiring duplicated unknowns or special enrichment functions.

Traditional fracture modeling approaches can be categorized into discrete crack methods, including extended FEMs [16-18], enrichments based on partition of unity [19-20], and RKPM near-tip enrichments [21-22], and diffuse crack methods, such as high-order gradient models [23-25] and phase field methods [26-29]. Discrete crack methods increase computational complexity by requiring crack detection and tracking. The diffuse crack approaches introduce regularization frameworks that yield a diffused representation of strong discontinuities, but high discretization resolution is usually required to achieve the desired accuracy. This work adopts a phase field model [30] for bulk brittle fracture and an interfacial nonlinear cohesive zone model [31] for interfacial debonding. The coupled two-field problem is further simplified to a single-field problem by introducing an energy-consistent strain-dependent damage model.

The remainder of this paper is organized as follows: Section 2 presents the basic equations for coupled bulk brittle fracture and interfacial decohesion modeling, along with the Reproducing Kernel Particle Method and numerical domain integration techniques. Section 3 details the SVM-guided image-based model construction, smeared interfacial displacement jump approximation, IM-RK approximation, and strain-dependent damage model. Section 4 showcases a numerical example of image-based modeling of microstructures, and Section 5 concludes with a discussion and summary.

## 2  Basic Equations

### 2.1  Smeared representation of cracked surface and material interfaces

Let $\Omega \in \mathbb{R}^d$ be an open domain with the space dimension $d$ describing a heterogeneous solid with an external boundary $\partial\Omega = \partial\Omega_g \cup \partial\Omega_h$, where $\partial\Omega_g$ and $\partial\Omega_h$ represent the Dirichlet and Neumann boundaries, respectively. We denote $\Gamma^C$ and $\Gamma^I$ as the crack surface and material interfaces. In this work, we utilize regularized functions in place of discontinuous functions on crack and interface surfaces $\Gamma^C$ and $\Gamma^I$ following the regularized framework proposed in [26-27]. Let the sharp crack topology be described by a time-dependent scalar auxiliary variable $d(\boldsymbol{x}, t)$ following the variational principle:

$$d(\boldsymbol{x}, t) = \text{Arg}\left\{\inf_{d \in \mathcal{S}^d} \Gamma^d(d, l_d)\right\}, \tag{1}$$

which subjected to the Dirichlet-type constraints: $\mathcal{S}^d = \{d | d(\boldsymbol{x}) = 1 \; \forall \boldsymbol{x} \in \Gamma\}$. Here, $l_d$ is a length-scale parameter, $\Gamma^d$ is a smeared representation of the total length of the sharp crack surface $\Gamma^C$, where

$$\Gamma^d = \int_\Omega \gamma_d(d, l_d) \, d\Omega. \tag{2}$$

In Eq. (2), $\gamma(d, l_d)$ denotes the crack surface density function per unit volume of the solid defined by [27]:

$$\gamma^d = \frac{1}{2l_d} d(\boldsymbol{x}, t)^2 + \frac{l_d}{2} \nabla d(\boldsymbol{x}, t) \cdot \nabla d(\boldsymbol{x}, t). \tag{3}$$

The regularization of the cracked surfaces is governed by the length-scale parameter $l_d$ such that for $l_d \to 0$, $\Gamma^d \to \Gamma^C$. Note that other crack surface density functions can be utilized, e.g., Borden et al. [32] proposed a smoother $\gamma^d$ with a fourth order Euler-Lagrange equation.

Similarly, the sharp material interfaces $\Gamma^I$ are regularized by a static variable $\beta(\boldsymbol{x})$ [30], which satisfies the Euler-Lagrange equation associated with the variational problem:

$$\beta(\boldsymbol{x}) = \text{Arg}\left\{\inf_{d \in \mathcal{S}^\beta} \Gamma^\beta(\beta, l_\beta)\right\}, \tag{4}$$

where $\mathcal{S}^\beta = \{\beta | \beta(\boldsymbol{x}) = 1 \; \forall \boldsymbol{x} \in \Gamma^I\}$ and $\Gamma^\beta = \int_\Omega \gamma_\beta(\beta, l_\beta) \, d\Omega$. $\Gamma^\beta$ represents the smeared total interface length with the interface density function $\gamma_\beta$ defined following Eq. (3):

$$\gamma^\beta = \frac{1}{2l_\beta}\beta(x)^2 + \frac{l_\beta}{2}\nabla\beta(x)\cdot\nabla\beta(x). \tag{5}$$

Note that the smeared material interfaces are also associated with a length-scale parameter. $l_\beta$, which is independent of the damage length-scale parameter $l_d$. The material interfaces are assumed to be fixed throughout the external loading stage so that unlike the auxiliary variable $d(x,t)$ associated with crack surfaces, $\beta(x)$ is time independent.

### 2.2 Potential Energy of Bulk Brittle Fracture and Cohesive Interfaces

With sharp discontinuities descriptions of both crack surface $\Gamma$ and material interfaces $\Gamma^I$, the potential energy functional is given by:

$$E = \int_{\bar\Omega\setminus\Gamma\cap\Gamma^I} W_u(\varepsilon(u))\,d\Omega + g_c\int_\Gamma d\Gamma + \int_{\Gamma^I} W^I(\llbracket u\rrbracket)\,d\Gamma - W^{ext}, \tag{6}$$

where $\varepsilon = \frac{1}{2}(\nabla u + (\nabla u)^T)$ is the infinitesimal strain tensor, and $W_u$, $g_c$, $W^I$ are the degraded elastic store energy density, bulk critical energy release rate, and strain energy density related to the interfacial displacement jump $\llbracket u \rrbracket$, respectively.

Since the bulk fracture and interface decohesion are accounted for separately, let $\varepsilon$ be split into $\varepsilon^e$ and $\tilde\varepsilon$, corresponding to the bulk behaviors and interfacial displacement jump, respectively [30], where $\tilde\varepsilon \to 0$ as $\beta(x) \to 0$. When the regularized crack and interface surface descriptions (Sec. 2.1) are adopted, Eq. (6) is replaced with:

$$E = \int_\Omega W(u,d,\beta)\,d\Omega - \int_\Omega u\cdot b\,d\Omega - \int_{\partial_h} u\cdot h\,d\Gamma, \tag{7}$$

where $b$ and $h$ are body force and applied traction, and $W$ is the free energy and is defined as:

$$W = W_u^e(\varepsilon^e(u,\beta),d) + [1-\beta(x)]g_c\gamma_d(d) + W^I(\llbracket u\rrbracket)\gamma_\beta(\beta). \tag{8}$$

As shown in Eq. (8), as $\beta(x) \to 1$, on the material interfaces, the dissipation functional is entirely governed by interface decohesion; when $\beta(x) \to 0$ thus $\gamma_\beta \to 0$ and $\varepsilon^e \to \varepsilon$, Eq. (7) recovers the regularized energy functional for brittle fracture proposed in [26-27].

Only the tension part of the stored energy is degraded [27], and $W_u$ is then defined as:

$$W_u(\varepsilon^e(u,\beta),d) = \psi_e^+(\varepsilon^e)[g(d)+\kappa] + \psi_e^-(\varepsilon^e). \tag{9}$$

$\psi_e^+$ and $\psi_e^-$ denote the tensile and compressive bulk strain energies in terms of principal bulk strain $\bar\varepsilon^e$:

$$\psi_e^{\pm} = \mu \langle \bar{\varepsilon}_i^e \rangle_{\pm} \langle \bar{\varepsilon}_i^e \rangle_{\pm} + \frac{\lambda}{2} \langle tr(\bar{\boldsymbol{\varepsilon}}^e) \rangle_{\pm}^2, \tag{10}$$

where the summation notation is adopted, and $\lambda$ and $\mu$ are Lamé's first and second parameters, respectively. Macaulay bracket notation is used in Eq. (10), where $\langle x \rangle_{\pm} = \frac{1}{2}(x \pm |x|)$. We adopt a quadratic degradation function $g(d) = (1-d)^2$ [27], and $\kappa \ll 1$ is introduced to maintain the well-posedness of the system for partially broken parts of the domain. Therefore, the displacement field $\boldsymbol{u}(\boldsymbol{x})$ can be found by minimizing the regularized global energy storage functional in Eq. (7):

$$\boldsymbol{u}(\boldsymbol{x}) = \text{Arg}\left\{ \inf_{\boldsymbol{u} \in \mathcal{S}^u} E(\boldsymbol{u}, d, \beta) \right\}, \tag{11}$$

where $\mathcal{S}^u = \{\boldsymbol{u} | \boldsymbol{u}(\boldsymbol{x}) = \boldsymbol{g} \; \forall \boldsymbol{x} \in \partial \Omega_g, \boldsymbol{u} \in H^1(\Omega)\}$.

## 2.3 Reproducing kernel (RK) approximation

Let a closed domain $\bar{\Omega} = \Omega \cup \partial \Omega \subset \mathbb{R}^d$ be discretized by a set of $NP$ nodes denoted by $\mathbb{S}^{\text{RK}} = \{\boldsymbol{x}_1, \boldsymbol{x}_2, \ldots, \boldsymbol{x}_{NP} \mid \boldsymbol{x}_I \in \bar{\Omega}\}$, as shown in Fig. 1, and let the approximation of a field variable $\boldsymbol{u}(\boldsymbol{x})$ in $\bar{\Omega}$ be denoted by $\boldsymbol{u}^h(\boldsymbol{x})$. The RK approximation of the field variable $\boldsymbol{u}(\boldsymbol{x})$ based on the discrete points in the set $\mathbb{S}^{\text{RK}}$ is formulated as follows:

$$u_i^h(\boldsymbol{x}) = \sum_{I=1}^{NP} \Psi_I(\boldsymbol{x}) d_{iI}, \tag{12}$$

where $\Psi_I$ denotes the RK shape function with support centered at the node $\boldsymbol{x}_I$ and $d_{iI}$ is the nodal coefficient in $i^{th}$ dimension to be sought. Moreover, let a node $I$ be associated with a subdomain $\Omega_I$, over which a kernel function $\phi_a(\boldsymbol{x} - \boldsymbol{x}_I)$ with a compact support $a$ is defined, such that $\bar{\Omega} \subset \cup_{I \in \mathbb{S}^{\text{RK}}} \Omega_I$ holds. The RK approximation function is constructed as:

$$\Psi_I(\boldsymbol{x}) = C(\boldsymbol{x}; \boldsymbol{x} - \boldsymbol{x}_I) \phi_a(\boldsymbol{x} - \boldsymbol{x}_I) = \left( \sum_{|\alpha| \leq n} (\boldsymbol{x} - \boldsymbol{x}_I)^{\alpha} b_{\alpha}(\boldsymbol{x}) \right) \phi_a(\boldsymbol{x} - \boldsymbol{x}_I) \tag{13}$$
$$\equiv \boldsymbol{H}^{\text{T}}(\boldsymbol{x} - \boldsymbol{x}_I) \boldsymbol{b}(\boldsymbol{x}) \phi_a(\boldsymbol{x} - \boldsymbol{x}_I),$$

$$\boldsymbol{H}^T(\boldsymbol{x} - \boldsymbol{x}_I) = [1, x_1 - x_{1I}, x_2 - x_{2I}, \ldots, (x_3 - x_{3I})^n], \tag{14}$$

where $\alpha$ is a multi-index notation such that $\alpha = (\alpha_1, \alpha_2, \ldots, \alpha_d)$ with a length defined as $|\alpha| = \alpha_1 + \alpha_2 + \cdots + \alpha_d$, and $\boldsymbol{x}^{\alpha} \equiv \boldsymbol{x}_1^{\alpha_1} \cdot \boldsymbol{x}_2^{\alpha_2}, \ldots, \boldsymbol{x}_d^{\alpha_d}$, $b_{\alpha} = b_{\alpha_1 \alpha_2 \cdots \alpha_d}$. The term $C(\boldsymbol{x}; \boldsymbol{x} - \boldsymbol{x}_I) = \boldsymbol{H}^{\text{T}}(\boldsymbol{x} - \boldsymbol{x}_I)\boldsymbol{b}(\boldsymbol{x})$ is called the correction function of the kernel $\phi_a(\boldsymbol{x} - \boldsymbol{x}_I)$ designed to introduced completeness to the RK approximation. The terms $\{(\boldsymbol{x} - \boldsymbol{x}_I)^{\alpha}\}_{|\alpha| \leq n}$ form a set of basis functions, and $\boldsymbol{H}^T(\boldsymbol{x} - \boldsymbol{x}_I)$ is the corresponding

vector of basis functions to the order $n$. The vector $\boldsymbol{b}(\boldsymbol{x})$ is the coefficient vector of $\{b_\alpha(\boldsymbol{x})\}_{|\alpha|\leq n}$ and is solved by enforcing the discrete reproducing conditions [33]:

$$\sum_{I=1}^{NP} \Psi_I(\boldsymbol{x})(\boldsymbol{x}-\boldsymbol{x}_I)^\alpha = \delta_{0\alpha}, \quad |\alpha| \leq n. \tag{15}$$

After inserting Eq. (13) into Eq. (15), $\boldsymbol{b}(\boldsymbol{x})$ is obtained as:

$$\boldsymbol{b}(\boldsymbol{x}) = \boldsymbol{M}^{-1}(\boldsymbol{x})\boldsymbol{H}(\boldsymbol{0})\phi_a(\boldsymbol{x}-\boldsymbol{x}_I), \tag{16}$$

where $\boldsymbol{M}(\boldsymbol{x})$ is the moment matrix and is formulated as:

$$\boldsymbol{M}(\boldsymbol{x}) = \sum_{I=1}^{NP} \boldsymbol{H}(\boldsymbol{x}-\boldsymbol{x}_I)\boldsymbol{H}^{\mathrm{T}}(\boldsymbol{x}-\boldsymbol{x}_I)\phi_a(\boldsymbol{x}-\boldsymbol{x}_I). \tag{17}$$

Finally, the RK shape function is obtained as:

$$\Psi_I(\boldsymbol{x}) = \boldsymbol{H}^{\mathrm{T}}(\boldsymbol{0})\boldsymbol{M}^{-1}(\boldsymbol{x})\boldsymbol{H}(\boldsymbol{x}-\boldsymbol{x}_I)\phi_a(\boldsymbol{x}-\boldsymbol{x}_I). \tag{18}$$

RK approximation allows separate controls over the order of completeness and the smoothness of approximation. Therefore, it can introduce high-order continuity into the approximation space, independent of the basis order, as the smoothness of the approximation functions is directly inherited from the smoothness of the kernel functions.

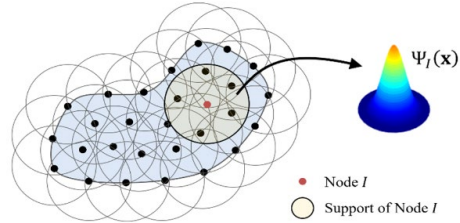

**Fig. 1.** Illustration of RK discretization and shape function

## 3    Image-based Modeling of Bulk and Interface Fractures

### 3.1    Image-based numerical model generation using the support vector machine

This work utilizes the support vector machine (SVM) to aid the material segmentation of micro-CT images [5-7, 15, 34-35]. A 30×30 pixels region of interest (ROI) containing four irregularly shaped inclusion particles (area contained in the red box in Fig. 2) is selected to demonstrate the training of SVM and numerical model generation processes, where the white areas in the sample image indicate the alumina inclusion material and the grey areas represent the epoxy material in the matrix. The training data are the physical coordinates of the centroids of pixel cells in the sample image, and the response labels are created by pre-processing the sample image using Otsu's method [36]. Note that only the pixel centroids' material classes and physical coordinates are provided as labeled training data, and training aims to identify material classes at arbitrary locations within the image domain, not limited to the image resolution.

Several hyperparameters must be determined beforehand to facilitate SVM's classification, including the kernel function, kernel scale, and misclassification penalty weight parameter. The Gaussian radial basis function is selected as the kernel function as inclusions are distinctive small particles. The selections of the kernel scale and penalty weight parameter are optimized utilizing an iterative Bayesian optimization process for 30 iterations, and the objective function of the Bayesian optimization process is to minimize the 5-fold cross-validation classification loss. The final training result for the ROI is shown in Fig. 3.

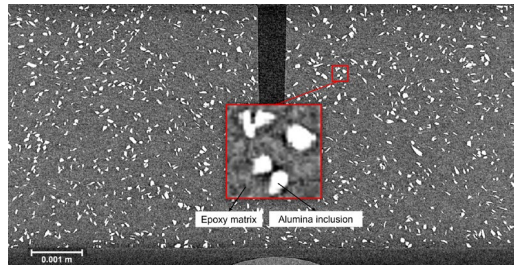

**Fig. 2.** Sample alumina-epoxy micro-CT image slice

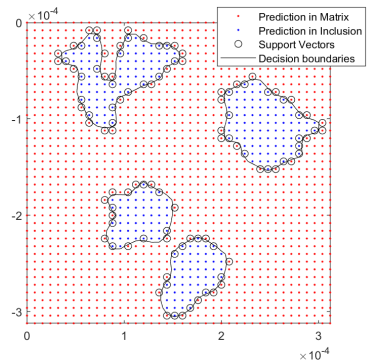

**Fig. 3.** SVM training results of the sample image

**Node-based numerical model generation**
As mentioned in Sec. 2.3, RK approximation utilizes node-based domain discretization. A set of uniformly distributed nodes $\mathbb{S}^0 \equiv \{x_I\}_{I=1}^{NP_0}$ is first constructed that coincide with SVM's training data points. In the interface-modified RK (IMRK) approximation, a set of interface nodes is generated [15] for the purpose of constructing weak and strong discontinuity across material interfaces, as shown in Fig. 4, where the material interfaces are represented by a simple line connection for visualization purposes.

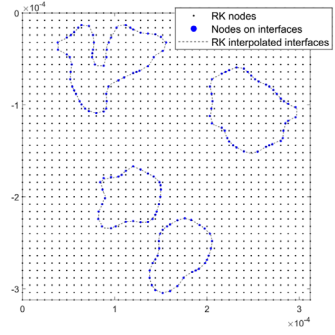

**Fig. 4.** SVM-RK numerical model for the test ROI

### 3.2  Interface-modified reproducing kernel (IM-RK) approximation

To enhance and localize the damage induced by interfacial decohesion, a weak discontinuity across the material interfaces is introduced by modifying the regular RK kernel function with a regularized Heaviside function $\widetilde{H}$ as follows [15]:

$$\bar{\phi}_a(\boldsymbol{x} - \boldsymbol{x}_I) = \phi_a(\boldsymbol{x} - \boldsymbol{x}_I)\widetilde{H}\left(\bar{\bar{\xi}}_I(\boldsymbol{x})\right), \tag{19}$$

where $\bar{\phi}_a(\boldsymbol{x} - \boldsymbol{x}_I)$ is a modified kernel function, and $\widetilde{H}(\cdot)$ and $\bar{\bar{\xi}}_I(\boldsymbol{x})$ in Eq. (19) are defined as:

$$\widetilde{H}(\cdot) = \max(0, \tanh(\cdot)), \tag{20}$$

and

$$\bar{\bar{\xi}}_I(\boldsymbol{x}) = \begin{cases} -\dfrac{S(\boldsymbol{x})}{c}, & S(\boldsymbol{x}_I) < 0 \\ +\dfrac{S(\boldsymbol{x})}{c}, & S(\boldsymbol{x}_I) > 0 \end{cases}, \tag{21}$$

where $c$ denotes a scaling factor with a length of the order of nodal spacing. Note that $S(\boldsymbol{x})$ is a signed distance of an evaluation point to its nearby interface, which is given from the output of the SVM-RK image segmentation and is readily available for evaluation of regularized Heaviside function $\widetilde{H}$. This normalized distance measure $\bar{\bar{\xi}}(\boldsymbol{x})$ is applicable to general n-dimensional image data. The kernel functions associated with nodes away from the interfaces have been scaled to zero at the material interfaces by the regularized Heaviside function, and the kernel functions associated with the interface nodes are not scaled.

The IM-RK shape functions are then constructed using the reproducing conditions given in Sec. 2.3. Fig. 5(a) and (b) show the IM-RK shape functions of non-interface nodes near the interfaces and the IM-RK shape functions of the interface nodes, constructed on the image-based SVM-RK discretized model shown in Fig. 4. The resulting IM-RK shape functions are truncated across arbitrarily shaped interfaces. However, the interface nodes' shape functions provide support coverage to the nodes located on both sides of the interface with $C^0$ continuity along the interfaces' normal direction for embedding weak discontinuities normal to the interface.

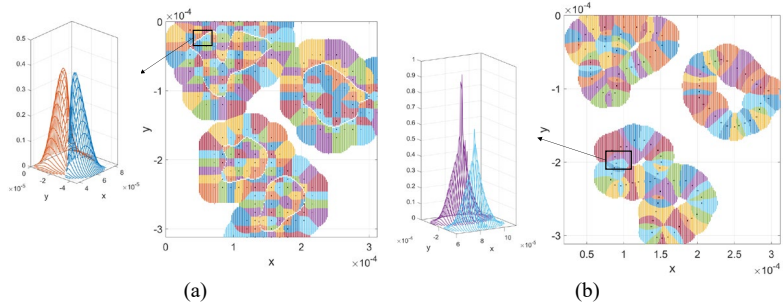

**Fig. 5.** IM-RK shape functions (a) for nodes around the interfaces and (b) for interface nodes: top view (right) and the zoom-in plots of two shape functions in the black box (left)

### 3.3 Smeared approximation of interfacial displacement jump

Instead of approximating the displacement field with strong discontinuities across material interfaces, we adopt a regularized smooth approximation for the displacement jump $[\![u]\!](x)$ similar to that in [30]. Recall that the material interfaces are identified by zero score function $S(x)$ obtained from SVM classification, and $S(x) > 0 \ \forall x \in \Omega^m$, $S(x) < 0 \ \forall x \in \Omega^i$, where $\Omega^i$ and $\Omega^m$ denote the inclusion and matrix domains, respectively. Then, the outward unit normal at a point $x^* \in \Gamma^I$ can be calculated as:

$$\vec{n}^I(x^*) = \frac{\nabla \tilde{S}(x)}{\|\nabla \tilde{S}(x)\|}\bigg|_{x=x^*} \tag{22}$$

Let $h \ll \Gamma^\beta$ and $x^{\pm*} = x^* \pm \frac{h}{2}\vec{n}^I(x^*)$ be the two points slightly perturbed from $x^*$ into the matrix and inclusion domains, as shown in Fig. 6. The displacement at $x^{\pm*}$ can be approximated using a first-order Taylor's expansion:

$$u(x^{\pm*}) = u\left(x^* \pm \frac{h}{2}\vec{n}^I(x^*)\right) \approx u(x^*) \pm \frac{h}{2}\nabla u(x^*) \cdot \vec{n}^I(x^*) \tag{23}$$

Since $S(x)$ is continuous in the domain $\Omega$, the calculation of $\vec{n}^I$ in Eq. (22) is valid for all $x \in \Omega$. Then the displacement jump evaluated at an arbitrary point $x \in \Omega$ can be approximated as:

$$[\![u]\!](x) \approx u(x^+) - u(x^-) = h\nabla u(x)\vec{n}^I(x) \coloneqq w(x) \tag{24}$$

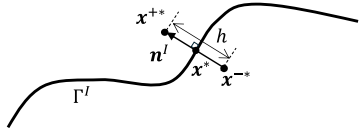

**Fig. 6.** Approximation of displacement jump across material interfaces

Substituting $[\![\boldsymbol{u}]\!](\boldsymbol{x})$ in Eq. (7)-(8) with the smeared interfacial displacement jump $\boldsymbol{w}(\boldsymbol{x})$, we obtain the weak form for $\boldsymbol{u}(\boldsymbol{x}) \in \mathcal{S}^u$ by finding the stationary of Eq. (11) with $\delta \boldsymbol{u}(\boldsymbol{x}) \in H_0^1$:

$$\int_\Omega \boldsymbol{\sigma}^e : \boldsymbol{\varepsilon}^e(\delta \boldsymbol{u}) \, d\Omega + \int_\Omega \gamma_\beta(\beta) \boldsymbol{t}^I(\boldsymbol{w}) \cdot \delta \boldsymbol{w} \, d\Omega - \int_\Omega \boldsymbol{b} \cdot \delta \boldsymbol{u} \, d\Omega \\ - \int_{\partial \Omega_h} \boldsymbol{h} \cdot \delta \boldsymbol{u} \, d\Gamma = 0, \qquad (25)$$

where $\boldsymbol{\sigma}^e = \frac{\partial W_u^e}{\partial \boldsymbol{\varepsilon}^e}$ is the Cauchy stress and $\boldsymbol{t}^I(\boldsymbol{w}) = \frac{\partial W^I}{\partial \boldsymbol{w}}$ is the traction vector associated with the interfacial displacement jump $\boldsymbol{w}$ acting on the interfaces oriented in $\vec{\boldsymbol{n}}^I$. Assume there is no body force and, by the balance of linear momentum and requiring the $\tilde{\boldsymbol{\varepsilon}}$ associated with the smoothed jump at interfaces not to exert external power, we have:

$$\int_\Omega \boldsymbol{\sigma}^e : \left[ \boldsymbol{\varepsilon}^e(\delta \boldsymbol{u}) + \vec{\boldsymbol{n}}^I \otimes^s \delta \boldsymbol{w} \gamma_\beta(\beta) - \nabla^s \delta \boldsymbol{u} \right] d\Omega = 0. \qquad (26)$$

Eq. (26) yields an admissible bulk strain field in the form:

$$\boldsymbol{\varepsilon}^e = \nabla^s \boldsymbol{u} - \vec{\boldsymbol{n}}^I \otimes^s \boldsymbol{w} \gamma_\beta(\beta), \qquad (27)$$

and $\tilde{\boldsymbol{\varepsilon}} = \vec{\boldsymbol{n}}^I \otimes^s \boldsymbol{w} \gamma_\beta(\beta)$.

### 3.4 Strain-dependent damage variable and interface cohesive model

Recall that the bulk material damage is governed by the auxiliary field (phase field) variable $d(\boldsymbol{x}, t)$. It can be solved used the thermodynamic force $\mathcal{A} = -\frac{\partial W}{\partial d}$ by enforcing the thermodynamic consistency [26-27, 30], which results in a coupled two-field problem where $d(\boldsymbol{x}, t)$ and $\boldsymbol{u}(\boldsymbol{x})$ are solved in a staggered manner. Alternatively, we assume $d$ is bulk strain dependent, i.e. $d = d(\boldsymbol{\varepsilon}^e, t)$. By enforcing the variational consistency ($\delta E = 0$) and using the expression of $\boldsymbol{\varepsilon}^e$ in Eq. (27), we get:

$$\frac{\partial g(d(\boldsymbol{\varepsilon}^e, t))}{\partial d} \psi_e^+ + (1 - \beta(\boldsymbol{x})) g_c \frac{\partial \gamma_d}{\partial d} = 0. \qquad (28)$$

Let $\gamma_d$ takes a simpler form than Eq. (3) that omits the higher order term $\mathcal{O}(\nabla d^2)$:

$$\gamma_d = \frac{1}{2l_d} d^2, \tag{29}$$

then, with the quadratic degradation function defined, Eq. (28) becomes an algebraic equation, and $d$ can be solved as:

$$d(\mathbf{x}, t) = \frac{2\psi_e^+}{2\psi_e^+ + \frac{(1-\beta(\mathbf{x}))g_c}{l_d}}. \tag{30}$$

To achieve the irreversibility of the damage, we introduce a tensile strain history function [26]:

$$\mathcal{H}(\mathbf{x}, t) = \max_{\tau \in [0,t]} \{\psi_e^+(\mathbf{x}, \tau)\}. \tag{31}$$

Replacing $\psi_e^+$ in Eq. (30) with Eq. (31), we get:

$$d(\mathbf{x}, t) = \frac{2\mathcal{H}(\mathbf{x}, t)}{2\mathcal{H}(\mathbf{x}, t) + \frac{(1-\beta(\mathbf{x}))g_c}{l_d}}, \tag{32}$$

and the displacement field is the sole unknown.

The interface damage is represented by an exponential cohesive zone mode (CZM) proposed by Xu and Needleman [31]. With the assumption of equal normal and tangential work of separation and zero coupling parameter. The potential of this CZM reads:

$$W^I(w_n, w_t) = g_c^I \left[1 - \left(1 + \frac{w_n}{\delta_n}\right) \exp\left(-\frac{w_n}{\delta_n}\right) \exp\left(-\frac{w_t^2}{\delta_t^2}\right)\right], \tag{33}$$

where $w^n = \mathbf{w} \cdot \overline{\mathbf{n}}^I$ and $w^t = \mathbf{w} \cdot \overline{\mathbf{m}}^I$ with $\overline{\mathbf{m}}^I$ being the tangential unit vector along interfaces, $g_c^I$ is the critical energy release rate associated with interfaces. $\delta_n$ and $\delta_t$ are characteristic lengths related to normal and tangential debonding, and:

$$\delta_n = \frac{g_c^I}{T_{n,max} \exp(1)}, \delta_t = \frac{g_c^I}{T_{t,max} \sqrt{0.5 \exp(1)}}, \tag{34}$$

where $T_{n,max}$ and $T_{t,max}$ are maximum normal and tangential tractions, respectively. Then the normal and tangential tractions along interfaces can be obtained as:

$$\begin{aligned} t_n^I &= \mathbf{t}^I \cdot \overline{\mathbf{n}}^I = \frac{\partial W^I}{\partial w_n} = \frac{g_c^I w_n}{\delta_n^2} \exp\left(-\frac{w_n}{\delta_n}\right) \exp\left(-\frac{w_t^2}{\delta_t^2}\right), \\ t_t^I &= \mathbf{t}^I \cdot \overline{\mathbf{m}}^I = \frac{\partial W^I}{\partial w_t} = \frac{2g_c^I w_t}{\delta_t^2} \left(1 + \frac{w_n}{\delta_n}\right) exp\left(-\frac{w_n}{\delta_n}\right) \exp\left(-\frac{w_t^2}{\delta_t^2}\right) \end{aligned} \tag{35}$$

## 4 Numerical Example

### 4.1 Damage modeling of three-point bending test with alumina-epoxy composite

A damage modeling of a three-point bending test of a single-edge notched alumina-epoxy composite specimen is considered [3]. Micro-CT images of the testing sample with a voxel size of 8 $\mu$m were taken after each loading step, and Fig. 7 (a) demonstrates a 2D slice of the composite beam with 5 vol% filler fractions. The sample height and thickness are 4.896 and 4.833 mm, respectively, and the loading span is 20 mm. The material properties for alumina inclusions and epoxy matrix are $E_i = 320$ GPa, $\nu_i = 0.23$, $E_m = 3.66$ GPa, $\nu_m = 0.358$. The critical energy release rates for the inclusions, matrix, and interfaces are $G_{ci} = 0.137$, $G_{cm} = 0.536$, and $G_{cI} = 0.0171$ N/mm, respectively. The length scale for displacement jump approximation is chosen as $h = 0.008$ mm, and the regularization length scales for bulk and interface damage are set as $l_d = l_\beta = 0.006$ mm. The bottom left corner and the bottom right corner of the beam are prescribed with the pin and roller boundary conditions and a $y$- directional displacement with a constant increment of $\bar{u}_y$ is prescribed to the center of the top edge. The rest of the domain boundaries are traction-free. Plane strain condition and pure mode I fracture are considered for this problem.

Since cracks were only observed in the region right above the notch, only the center region with a width of 0.896 mm is modeled with detailed microstructures (Sec. 3.1), and the rest of the beam is modeled as a homogenized material with a weighted average material property:

$$X_h = (1 - p)X_m + pX_i, \tag{36}$$

where $p$ is the volume fraction of alumina inclusions, and $X$ can be $E$, $\nu$, or $g_c$ with the subscript "$h$" denotes the homogenized properties. Additionally, a blending zone (brown shaded zones in Fig. 7. (b)) of 0.32 mm is defined between the heterogeneous and homogenized regions, in which the material properties are blended by:

$$X_b = (1 - \alpha(x))X_{micro} + \alpha(x)X_h, \tag{37}$$

where $\alpha(x)$ is taken as a scaled sigmoid function that equals 1 at the homogenized domain boundaries and 0 at the microstructure boundaries.

Fig. 8 shows the crack patterns observed with micro-CT imaging and those obtained from the numerical simulation at different prescribed displacements. Cracks initiate around the material interfaces and then propagate into the matrix from the corners of inclusions. Crack coalesces around closely positioned inclusions are also observed in both experimental and numerical results. In addition, Fig. 9 shows a comparison of the load-displacement curves between the mechanical testing with and without micro-CT imaging and the numerical simulation. The experiment with in-situ micro-CT imaging needs to relax the sample after each loading step [3], leading to multiple jumps in the load-displacement curves. The numerical simulation's load-displacement behavior agrees well with the experimental one, both in the initial slope and the peak load.

Overall, the numerical simulation results are in good agreement with the experimental observations.

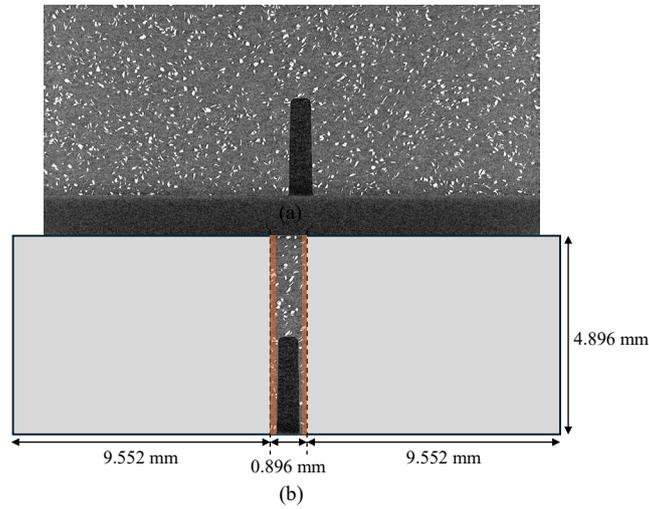

**Fig. 7.** (a) Micro-CT slice of the alumina-epoxy composite beam and (b) regions for homogenized (light gray) and detailed microstructure modeling

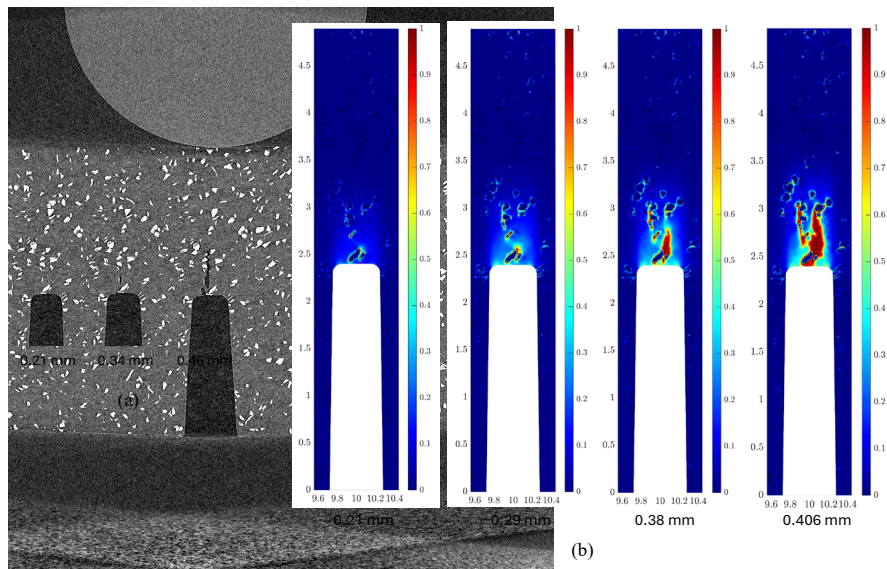

**Fig. 8.** Crack patterns at different prescribed displacements: (a) experimental results; (b) numerical simulation results

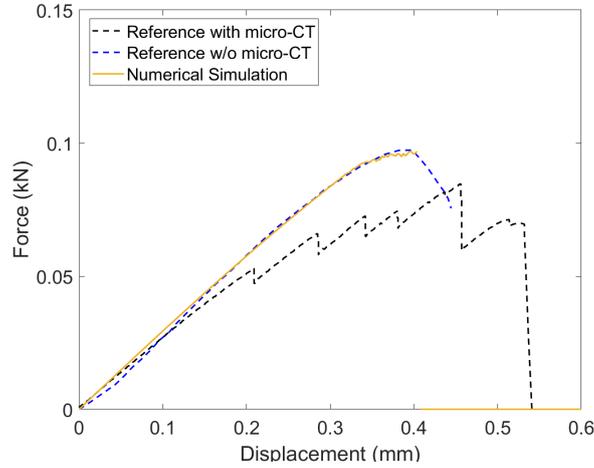

**Fig. 9.** Comparison of load-displacement curves obtained from numerical simulation and experiments with and without in-situ micro-CT imaging

## 5 Conclusion

This work introduces a phase field-based damage model coupled with an interface cohesive zone model to simulate interactions between bulk and interface cracks in composite materials with arbitrary inclusion geometries. We propose an image-based modeling guided by Support Vector Machine (SVM) using micro-CT images of composite materials. The SVM classification scores, representing signed distances to the material interfaces, enable the identification of material phases, interface discretization, and interface surface normals, facilitating the automatic construction of RK discretization and approximation with weak discontinuities. The resulting Interface-Modified Reproducing Kernel Particle Method (IM-RKPM) properly captures the damage-induced localizations. The effectiveness of this framework is demonstrated by a numerical simulation of a three-point bending test of an epoxy-alumina composite structure. Future work will incorporate neural network (NN) enrichments into the RK approximations, where the location, orientation, and regularization widths of the evolving localization paths will be automatically captured by the machined learned NN parameters.